\documentclass{llncs}
\usepackage{amsfonts}
\usepackage{amscd}
\usepackage{amssymb}
\usepackage{pifont}
\usepackage{epsfig}
\begin{document}
\title{\bf Quantum Authentication and Quantum Key Distribution Protocol}

\author{Hwayean Lee \inst{1,2,3}, Jongin Lim\inst{1,2}, and HyungJin Yang\inst{2,4}} \institute{
Center for Information Security Technologies(CIST)\inst{1},\\
Korea University, Anam Dong, Sungbuk Gu, Seoul, Korea\\
Graduate School of Information Security(GSIS) \inst{2}\\
Institut f\"{u}r Experimentalphysik, Universit\"{a}t Wien, Austria
\inst{3}\\ Department of Physics, Korea University, Chochiwon,Choongnam, Korea \inst{4} \\
\email{ \{hylee, jilim, yangh\}@korea.ac.kr }}
%\date{}
\maketitle

\begin{abstract}
We propose a quantum key distribution protocol with quantum based
user authentication. Our protocol is the first one in which users
can authenticate each other without previously shared secret and
then securely distribute a key where the key may not be exposed to
even a trusted third party. The security of our protocol is
guaranteed by the properties of the entanglement.
\end{abstract}

\section{Introduction}
Quantum key distribution(QKD) is the most actively researched field
in Quantum Cryptography. Since BB84 protocol\cite{BB84} was proposed
by Bennett and Brassard in 1984 as a start, many QKD protocols have
been proposed\cite{Eke91,b92,BBM} and
implemented\cite{JZ00,HMP00,GZ02}. The great advantage of QKD is to
provide the provable security of distributed
keys\cite{SP00,M98,LC99}. However, it is assumed that the quantum
channel is directly connected and previously authorized to the
designated users in those protocols. This assumption is not suitable
on the consideration of quantum networks. To authenticate users on
the quantum networks, Quantum Authentication
protocols\cite{CS01,SG01,CF02,CS95,DM99,ZZ00,M02,LK00} are proposed
since Crepeau and L. Salvail first proposed a quantum identification
protocol in 1995. Some Quantum Authentication protocols assume that
the users have some authentication information such as entangled
states\cite{CS01,SG01,CF02} and authentication
sequence\cite{CS95,DM99}. As mentioned above, these protocols can
not be operated on the quantum networks. Other quantum
authentication protocols\cite{ZZ00,M02,LK00} introduced a trusted
third party. Quantum authentication protocols proposed by Zeng and
Zhang\cite{ZZ00} in 2000 and Mihara\cite{M02} in 2002 are only for
authentication. Alice and Bob can authenticate each other and
distribute key without previously shared information only in one
protocol proposed by Ljunggren and et al.\cite{LK00}. The major
disadvantage of this protocol is the leakage of the key to the
trusted third party.

In this paper, we propose a Quantum Key Distribution protocol with
authentication. The proper users, Alice and Bob can authenticate
each other without previously shared secret and share a secret key
without leakage of information to anyone. We organize this paper as
follows. First, we propose a new QKD protocol with user
authentication in chapter 2. Our QKD protocol is composed of two
parts: one is authentication and the other key is distribution.
Greenberger - Horne - Zeilinger (GHZ) states\cite{GHZ} are used to
authenticate users and distribute a secret key. The security
analysis of our protocol is discussed in chapter 3 and at last our
conclusion is presented in chapter 4.

\section{Quantum Authentication and Quantum Key Distribution protocol}
\subsection{Authentication}
We assume that Alice and Bob do not share any prior secret
information or entanglement states for authentication. To identify
each other in the communication, they are supposed to introduce a
trusted third party, Trent. Trent plays a role like a CA(certificate
authority) in PKI(Public Key Infrastructure)\cite{stinson,PKI}. If
there are $n$ users in quantum networks, then $\frac{n(n-1)}{2}$
keys are needed to communicate freely when there is no Trent.
Besides, each user must distribute $n-1$ secret keys with other
users. However, only $n$ keys are needed when Trent exists and each
user just needs to distribute one secret key with Trent. Trent may
be a loophole for security. However it can be overcome using similar
methods applied to CA.

We assume that Alice has registered her secret identity $ID_A$ and a
one-way hash function $h_A : \{0,1\}^* \times \{0,1\}^l \rightarrow
\{0,1\}^m$, where $*$ means an arbitrary length, $l$ is the length
of a counter, and $m$ is a constant. Bob has also registered his
secret identity $ID_B$ and a one-way hash function $h_B$ to Trent.
This information is assumed to be kept secret between the user and
Trent. Authentication key can, then, be generated by a hashed value
$h_{user} (ID_{user} , c_{user} )$ where $c_{user}$ is a counter
which is the number of the calls of the one way hash function
$h_{user}$.

If Alice wants to distribute a key with Bob, she notifies this fact
to Bob and Trent. On receiving the request, Trent generates $N$ GHZ
tripartite states $\vert \Psi \rangle = \vert \psi_1 \rangle \vert
\psi_2 \rangle ... \vert \psi_N \rangle $. For simplicity the
following GHZ state $\vert \psi_i \rangle $ is supposed to be
prepared.
$$\vert \psi_i \rangle = \frac{1}{\sqrt{2}}(\vert 000 \rangle_{ATB}
+\vert 111 \rangle_{ATB} )$$ where the subscripts A, T and B
correspond to Alice, Trent, and Bob, respectively. In this paper, we
represent the z basis as $\{ \vert 0 \rangle , \vert 1 \rangle \}$
and the x basis as $\{ \vert + \rangle , \vert - \rangle \}$, where
$ \vert + \rangle = \frac{1}{\sqrt{2}} ( \vert 0 \rangle + \vert 1
\rangle  )$ and $ \vert - \rangle = \frac{1}{\sqrt{2}} ( \vert 0
\rangle - \vert 1 \rangle  )$.

Next, Trent encodes Alice's and Bob's particles of GHZ states with
their authentication keys, $h_A (ID_A , c_A )$ and $h_B (ID_B , c_B
)$, respectively. For example, if the $i$th value of $h_A (ID_A ,
c_A)$ is 0, then Trent makes an identity operation $I$ to Alice's
particle of the $i$th GHZ state. If it is 1, Hadamard operation $H$
is applied. If the authentication key does not have enough length to
cover all GHZ particles, new authentication keys can be created by
increasing the counter until the authentication keys shield all GHZ
particles. After making operations on the GHZ particles, Trent
distributes the states to Alice and Bob and keeps the remaining for
him.

\begin{figure}[h]
\begin{center}
\epsfig{figure=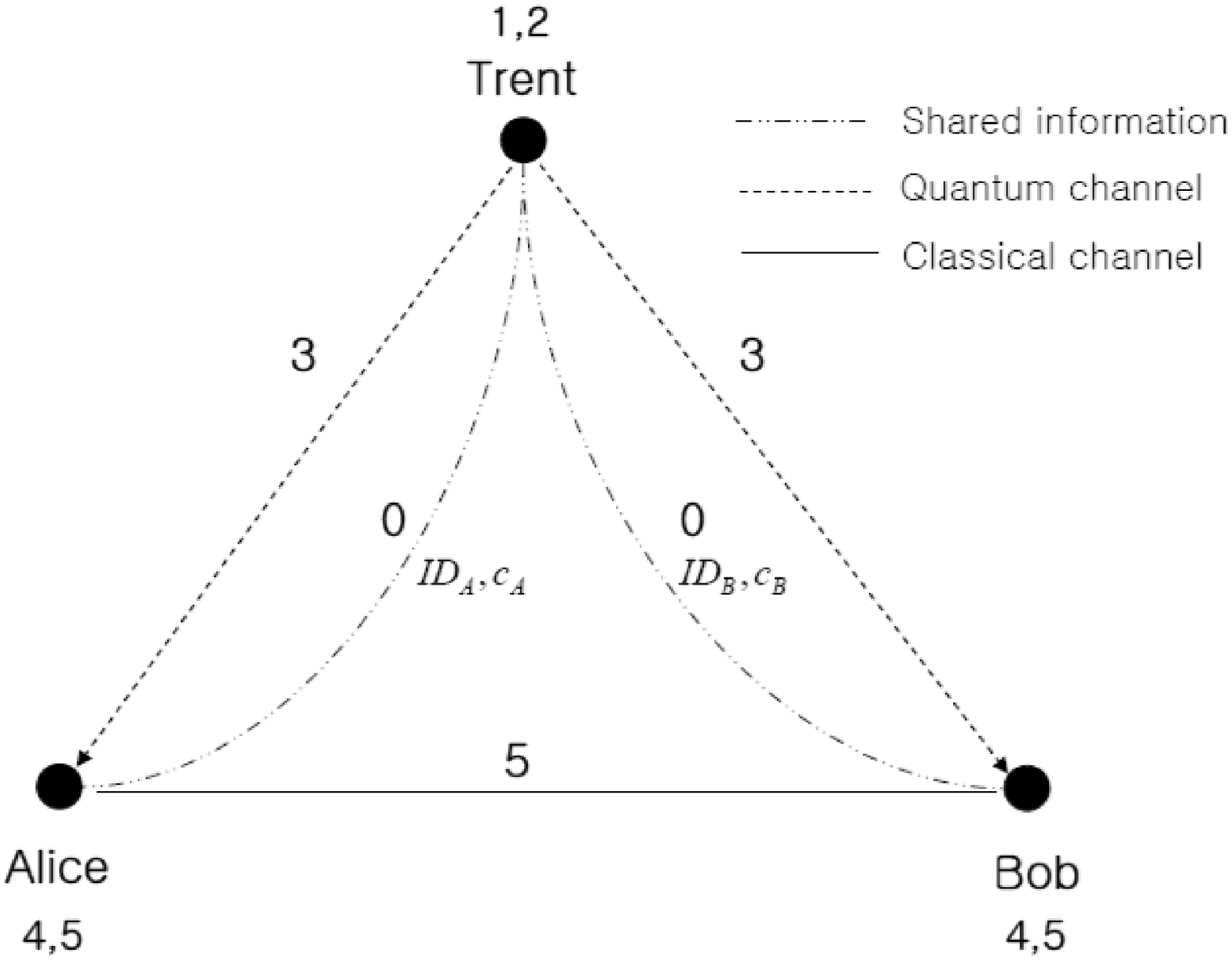, width=7cm, height=5.5cm}
\caption{\textbf{Procedures of Authentication} 0. Alice and Bob
register their secret identities and hash functions to Trent. 1.
Trent generates GHZ states $\vert \psi \rangle =
\frac{1}{\sqrt{2}}(\vert 000\rangle_{ATB}+\vert 111\rangle_{ATB})$.
2. Trent makes unitary operations on $\vert \psi \rangle$ with
Alice's and Bob's authentication key. 3. Trent distributes GHZ
particles to Alice and Bob. 4. Alice and Bob make reverse unitary
operations on their qubits with their authentication key,
respectively. 5. Alice and Bob choose the position of a subset of
GHZ states and make a local measurement in the $z$ basis on them and
compare the results.}\label{Aut}
\end{center}
\end{figure}

On receiving the qubits, Alice and Bob make reverse unitary
operations on their qubits with their authentication key $h_A (ID_A
, c_A )$ and $h_B (ID_B , c_B )$, respectively. This authentication
procedure can be written in the following form of sequences of local
unitary operation, the initial state:
\begin{displaymath} \vert \psi_i \rangle_{1}~ = ~
\frac{1}{\sqrt{2}}(\vert 000 \rangle_{ATB} +\vert 111 \rangle_{ATB}
) \end{displaymath} state after Trent's transformation
\begin{displaymath} \vert \psi_i \rangle_{2} ~ = ~ \{ [1-h_A (ID_A ,
c_A) ]I + [h_A (ID_A , c_A)] H \}_A \end{displaymath}
\begin{displaymath}  ~~~~~~~~~~~~~~~ \otimes \{ [1-h_B (ID_B , c_B) ]I
+ [h_B (ID_B , c_B)] H \}_B \vert \psi_i \rangle_{1}
\end{displaymath}
and finally the state after Alice's and Bob's local operations
\begin{displaymath}  \vert \psi_i \rangle_{3} ~= ~ \{ [1-h_A (ID_A ,
c_A) ]I + [h_A (ID_A  c_A)] H \}_A  \end{displaymath}
\begin{displaymath} ~~~~~~~~~~~~~~ \otimes \{ [1-h_B (ID_B , c_B) ]I +
[h_B (ID_B , c_B)] H \}_B \vert \psi_i \rangle_{2} \end{displaymath}
\begin{displaymath} = ~ \vert \psi_i \rangle_{1} \end{displaymath}
where $ \vert \psi_i \rangle $ is the state of the $i$-th GHZ
particle and the subscript 1, 2, and 3 represents the three steps of
authentication.

Next, Alice and Bob select some of the decoded qubits, make
von-Neumann measurements on them, and compare the results through
the public channel. If the error rate is higher than expected, then
Alice and Bob abort the protocol. Otherwise they can confirm that
the other party is legitimate and the channel is secure. They then
execute the following key distribution procedures.

\subsection{Key Distribution}
Alice and Bob randomly make an operation either identity operation
$I$ or Hadamard operation $H$ on the remaining GHZ particles. They
keep the record of the operations which they made. For example, 0
represents $I$ and 1 indicates $H$. After making unitary operations,
Bob sends his encrypted GHZ particles to Alice. On receiving the
qubits, Alice makes Bell measurements on pairs of particles
consisting of her qubit and Bob's qubit. On the other hand, Trent
measures his third qubit in the x basis and reveals the measurement
outcomes. In this paper we use the following notations of Bell
states.
$$\vert \Phi^+ \rangle = \frac{1}{\sqrt{2}} \{\vert 00 \rangle +\vert 11 \rangle
\}$$
$$\vert \Phi^- \rangle = \frac{1}{\sqrt{2}} \{\vert 00 \rangle - \vert 11 \rangle
\}$$
$$\vert \Psi^+ \rangle = \frac{1}{\sqrt{2}} \{\vert 01 \rangle + \vert 10 \rangle
\}$$
$$\vert \Psi^- \rangle = \frac{1}{\sqrt{2}} \{\vert 01 \rangle - \vert 10 \rangle
\}$$
\begin{figure}[h]
\begin{center}
\epsfig{figure=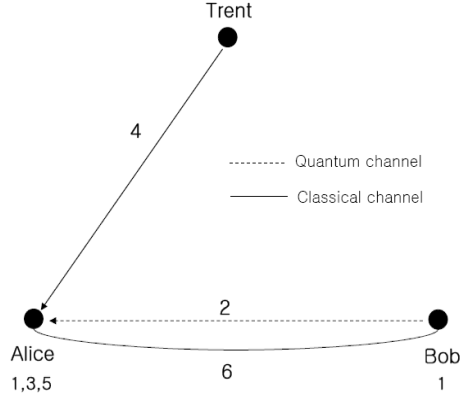, width=6.5cm, height=5.5cm}
\caption{\textbf{Procedures of Key distribution} 1. Alice and Bob
make identity operations I (0) or Hadamard operations H (1) randomly
on the remaining GHZ particles after authentication. 2. Bob sends
his encoded particles to Alice. 3. Alice makes Bell measurements on
pairs of particles consisting of her qubit and Bob's qubit. 4. The
arbitrator measures his qubits in the x basis and publishes the
results. 5. Alice infers Bob's operation using the table
[\ref{GHZ}]. 6. Alice and Bob select check bits and compare
them.}\label{KD}
\end{center}
\end{figure}

Alice can infer Bob's unitary operations and sometimes discover the
existence of Eve using the table [\ref{GHZ}]. For example, if Trent
discloses $\vert + \rangle$, Alice chooses $I$ operation and her
Bell measurement result is $\vert \Phi^- \rangle$, then Alice can
infer that Bob made a $H$ operation and he sent 1. On the other
hand, if Trent makes public $\vert + \rangle$, Alice makes $I$
operation and obtains $\vert \Psi^- \rangle$, then Alice can detect
an error.

\begin{table}[h]
\begin{center}
\caption{ {\small Operations on reversed GHZ states(i.e. $\vert \psi
\rangle $) and published information }}
\begin{tabular}{|c|c|c| } \hline
\multicolumn{2}{|c|} {Operation} & Transformation of GHZ states
\\\cline{1-2}   Alice  & Bob & after Alice's and Bob's operations \\ \hline

$I(0)$ & $I(0)$ & $\frac{1}{\sqrt{2}} (\vert \Phi^+ \rangle_{AB}
\vert + \rangle_a  + \vert \Phi^- \rangle_{AB}  \vert - \rangle_a ) $ \\
\hline

$I(0)$ & $H(1)$ & $\frac{1}{2} (\vert \Phi^+ \rangle_{AB} \vert -
\rangle_a  + \vert \Phi^- \rangle_{AB} \vert + \rangle_a + \vert
\Psi^+ \rangle_{AB} \vert + \rangle_a  + \vert \Psi^- \rangle_{AB}
\vert - \rangle_a ) $ \\ \hline

$H(1) $ & $I(0)$ & $\frac{1}{2} (\vert \Phi^+ \rangle_{AB} \vert -
\rangle_a  + \vert \Phi^- \rangle_{AB} \vert + \rangle_a + \vert
\Psi^+ \rangle_{AB} \vert + \rangle_a  - \vert \Psi^- \rangle_{AB}
\vert - \rangle_a ) $  \\ \hline

$H(1)$ & $H(1)$ & $\frac{1}{\sqrt{2}} (\vert \Phi^+ \rangle_{AB}
\vert + \rangle_a  + \vert \Psi^+ \rangle_{AB}  \vert - \rangle_a )$ \\
\hline
\end{tabular} \label{GHZ}
\end{center}
\end{table}

Alice and Bob compare some bits of their shared key (Bob's operation
sequence). If the error rate is higher than the acceptable level,
they throw away the shared sequence and restart the protocol.
Otherwise they use the remaining sequences as a secret key. Usual
error correction can be implemented to correct the remaining errors.
Alice and Bob can reduce the Eve's knowledge of a shared key by
standard privacy amplification\cite{CM97,GRTZ}.

\section{Security Analysis}
In the assumption, user identity and a hash function are enrolled to
Trent and the information is kept secret only between the owners and
the arbitrator. Moreover Trent is supposed to be a honest person
whom Alice and Bob can trust.

We first analyze the process of authentication. Suppose Eve
intercepts the qubits heading to Alice or Bob and disguises her or
him. Let Eve use the following unitary operation $U_{AE}$ on Alice's
and her qubit $\vert e \rangle$.
$$U_{AE} \vert 0e \rangle_{AE} = \alpha \vert 0\rangle_A \vert e_{00}
\rangle_E + \beta \vert 1\rangle_A \vert e_{01} \rangle_E $$
$$U_{AE} \vert 1e \rangle_{AE} = \beta' \vert 0 \rangle_A \vert e_{10}
\rangle_E + \alpha' \vert 1\rangle_A \vert e_{11} \rangle_E $$ where
$|\alpha|^2 + |\beta|^2 =1$, $|\alpha'|^2 + |\beta'|^2 =1$ and
$\alpha \beta^* + \alpha'^{*} \beta' =0$. If a bit of Alice's
authentication key is 0 (1), the total states $\vert \xi_0 \rangle$
(or $\vert \xi_1 \rangle$) of system and Eve's probe after Alice's
and Bob's reverse operation is as follows.
$$\vert \xi_0 \rangle = U_{AE} \{ \frac{1}{\sqrt{2}}
(\vert 000 \rangle_{ATB} +\vert 111 \rangle_{ATB}) \} \vert e
\rangle_E $$
$$=\frac{1}{\sqrt{2}} \{ \alpha \vert 000 \rangle_{ATB} \vert e_{00} \rangle_E
+ \beta \vert 100 \rangle_{ATB} \vert e_{01} \rangle_E $$
$$~~~+ \beta' \vert 011 \rangle_{ATB} \vert e_{10} \rangle_E + \alpha' \vert 111
\rangle_{ATB} \vert e_{11} \rangle_E \} $$
$$\vert \xi_1 \rangle =H_A  U_{AE} \{ H_A \frac{1}{\sqrt{2}}
(\vert 000 \rangle_{ATB} +\vert 111 \rangle_{ATB}) \} \vert
e\rangle_E $$
%$$=H_A  U_{AE} \{ \frac{1}{2}
%(\vert 000 \rangle_{ATB}+ \vert 100 \rangle_{ATB}+ \vert 011
%\rangle_{ATB} - \vert 111 \rangle_{ATB}) \} \vert e\rangle_E $$
%$$=H_A  \frac{1}{2} \{ (\vert 00 \rangle + \vert 11 \rangle)_{TB}
%(\alpha \vert 0 \rangle_A \vert e_{00} \rangle_E +\beta \vert 1
%\rangle_A \vert e_{01} \rangle_E)$$
%$$ +(\vert 00 \rangle - \vert 11
%\rangle)_{TB} (\beta' \vert 0 \rangle_A \vert e_{10} \rangle_E
%+\alpha' \vert 1 \rangle_A \vert e_{11} \rangle_E)\}$$
$$=\frac{1}{2\sqrt{2}} \{\vert000 \rangle_{ATB} (\alpha \vert e_{00}
\rangle_E + \beta \vert e_{01} \rangle_E + \beta' \vert e_{10}
\rangle_E + \alpha' \vert e_{11} \rangle_E )  $$
$$+\vert001 \rangle_{ATB} (\alpha \vert e_{00}
\rangle_E - \beta \vert e_{01} \rangle_E + \beta' \vert e_{10}
\rangle_E - \alpha' \vert e_{11} \rangle_E )$$
$$+\vert110 \rangle_{ATB} (\alpha \vert e_{00}
\rangle_E + \beta \vert e_{01} \rangle_E - \beta' \vert e_{10}
\rangle_E - \alpha' \vert e_{11} \rangle_E ) $$
$$+\vert111 \rangle_{ATB} (\alpha \vert e_{00}
\rangle_E - \beta \vert e_{01} \rangle_E - \beta' \vert e_{10}
\rangle_E + \alpha' \vert e_{11} \rangle_E )  \}$$

Eve can be detected with probability $\frac{1+\beta^2 +\beta'^2
}{4}$(when the probability of 0 and 1 in an authentication key is
same) in the authentication phase. If the number of the check bits
in the authentication process is $c$, then Alice and Bob can find
out the existence of Eve with probability of $1 - (\frac{1+\alpha^2
+\alpha'^2 }{4} )^c$. Eve is, therefore, always revealed if $c$ is
large enough. Hence if the authentication is passed, then Alice and
Bob confirm the other party is the designated user.

Moreover, the original secret identities of users cannot be revealed
even if Eve estimates some bits of the authentication key i.e. the
hashed value. Eve can infer only some bits of the authentication key
by checking bits in the authentication process. However Eve cannot
reverse the hash function with partial information of the hashed
value obtained from the checking bits in the authentication process.
Besides Eve cannot infer the next authentication key since it is
used only once and changed every time.

After authentication process, only Bob's qubits are transmitted. Eve
will make operations on these qubits in key distribution phase.
Suppose Eve use the above unitary operation $U_{BE}$ on Bob's and
her qubit $\vert E \rangle$. Then we can get the following states of
total system composed by Alice, Bob, Trent and Eve. Equation (1) is
derived from the situation when Alice and Bob choose $I$, equation
(2) when they apply different unitary operations($H$ and $I$), and
equation (3) is when they make $H$ operations.

\begin{itemize}
\item[(1)] $\frac{1}{2\sqrt{2}} \Big[ \vert \Phi^+ \rangle_{AB}
\big\{ \vert + \rangle_T (\alpha \vert e_{00}\rangle_E + \alpha'
\vert e_{11}\rangle_E ) + \vert - \rangle_T (\alpha \vert
e_{00}\rangle_E - \alpha' \vert e_{11}\rangle_E ) \big\} $

$ ~~~+ \vert \Phi^- \rangle_{AB} \big\{ \vert + \rangle_T (\alpha
\vert e_{00}\rangle_E - \alpha' \vert e_{11}\rangle_E ) + \vert -
\rangle_T (\alpha \vert e_{00}\rangle_E + \alpha' \vert
e_{11}\rangle_E )\big\}$

$~~~ + \vert \Psi^+ \rangle_{AB} \big\{ \vert + \rangle_T (\beta
\vert e_{01}\rangle_E + \beta' \vert e_{10}\rangle_E ) + \vert -
\rangle_T (\beta \vert e_{01}\rangle_E - \beta' \vert
e_{10}\rangle_E )\big\}$

$~~~ + \vert \Psi^- \rangle_{AB} \big\{ \vert + \rangle_T (\beta
\vert e_{01}\rangle_E - \beta' \vert e_{10}\rangle_E ) + \vert -
\rangle_T (\beta \vert e_{01}\rangle_E + \beta' \vert
e_{10}\rangle_E )\big\} \Big]$

\item[(2)]$\frac{1}{4} \Big[ \vert \Phi^+ \rangle_{AB}
\big\{ \vert + \rangle_T (\alpha \vert e_{00}\rangle_E - \alpha'
\vert e_{11}\rangle_E + \beta \vert e_{01}\rangle_E + \beta' \vert
e_{10}\rangle_E )$

$~~~~~~~~~~+ \vert - \rangle_T (\alpha \vert e_{00}\rangle_E +
\alpha' \vert e_{11}\rangle_E \mp \beta \vert e_{01}\rangle_E \pm
\beta' \vert e_{10}\rangle_E  ) \big\} $

$ + \vert \Phi^- \rangle_{AB} \big\{ \vert + \rangle_T (\alpha \vert
e_{00}\rangle_E + \alpha' \vert e_{11}\rangle_E - \beta \vert
e_{01}\rangle_E + \beta' \vert e_{10}\rangle_E  )$

$~~~~~~~~~~ + \vert - \rangle_T (\alpha \vert e_{00}\rangle_E -
\alpha' \vert e_{11}\rangle_E \pm \beta \vert e_{01}\rangle_E \pm
\beta' \vert e_{10}\rangle_E )\big\}$

$ + \vert \Psi^+ \rangle_{AB} \big\{ \vert + \rangle_T ( \alpha
\vert e_{00}\rangle_E + \alpha' \vert e_{11}\rangle_E + \beta \vert
e_{01}\rangle_E - \beta' \vert e_{10}\rangle_E ) $

$~~~~~~~~~~+ \vert - \rangle_T (\mp \alpha \vert e_{00}\rangle_E \pm
\alpha' \vert e_{11}\rangle_E  + \beta \vert e_{01}\rangle_E +\beta'
\vert e_{10}\rangle_E )\big\}$

$ + \vert \Psi^- \rangle_{AB} \big\{ \vert + \rangle_T (- \alpha
\vert e_{00}\rangle_E + \alpha' \vert e_{11}\rangle_E  + \beta \vert
e_{01}\rangle_E + \beta' \vert e_{10}\rangle_E )$

$~~~~~~~~~~ + \vert - \rangle_T (\pm\alpha \vert e_{00}\rangle_E \pm
\alpha' \vert e_{11}\rangle_E  + \beta \vert e_{01}\rangle_E -
\beta' \vert e_{10}\rangle_E )\big\} \Big]$

\item[(3)] $\frac{1}{2\sqrt{2}} \Big[ \vert \Phi^+ \rangle_{AB}
\big\{ \vert + \rangle_T (\alpha \vert e_{00}\rangle_E + \alpha'
\vert e_{11}\rangle_E ) + \vert - \rangle_T (\beta \vert
e_{01}\rangle_E + \beta' \vert e_{10}\rangle_E ) \big\} $

$ ~~~+ \vert \Phi^- \rangle_{AB} \big\{ \vert + \rangle_T (\alpha
\vert e_{00}\rangle_E - \alpha' \vert e_{11}\rangle_E ) - \vert -
\rangle_T (\beta \vert e_{01}\rangle_E - \beta' \vert
e_{10}\rangle_E )\big\}$

$~~~ + \vert \Psi^+ \rangle_{AB} \big\{ \vert + \rangle_T (\beta
\vert e_{01}\rangle_E + \beta' \vert e_{10}\rangle_E ) + \vert -
\rangle_T (\alpha \vert e_{00}\rangle_E + \alpha' \vert
e_{11}\rangle_E ) \big\}$

$~~~ + \vert \Psi^- \rangle_{AB} \big\{ \vert + \rangle_T (\beta
\vert e_{01}\rangle_E - \beta' \vert e_{10}\rangle_E ) - \vert -
\rangle_T (\alpha \vert e_{00}\rangle_E - \alpha' \vert
e_{11}\rangle_E ) \big\} \Big]$
\end{itemize}

As shown in the above equations, Eve can be detected with
probability $\frac{1}{2} + \frac{\beta^2 +\beta'^2 }{8}$ per check
bit in the key distribution phase. Hence Eve can be detected with
certainly if enough check bits are used in the key distribution. In
this regard, Alice and Bob can identify and securely distribute a
key with certainty using our protocol.

\section{Conclusions}
We propose a quantum key distribution protocol with quantum based
user authentication. User authentication is executed without
previously shared secret and by validating the correlation of GHZ
states. A key can be securely distributed by using the remaining GHZ
states after authentication. By the properties of the entanglement
of GHZ states, even the trusted third party, Trent can not get out
the distributed key. We expect our protocol can well be adjusted to
be incorporated in future quantum networks.

We acknowledge helpful discussion with Andreas Poppe and Hannes
H\"{u}bel. This work was supported by the Korea Research Foundation
Grant funded by the Korean Government(MOEHRD)(KRF-2005-213-D00090).

\end{document}